\documentclass[aps,prl,reprint,groupedaddress,amsmath,amssymb,longbibliography]{revtex4-1}
\usepackage{amsmath}
\usepackage{amssymb}
\usepackage{amsthm}
\usepackage{amsfonts}
\usepackage{subfigure}
\usepackage{graphicx}
\usepackage{dcolumn}
\usepackage{bm}

\usepackage{booktabs, array, mathptmx, float, tabularx, booktabs, lipsum, multirow}
\usepackage{siunitx, xcolor}
\usepackage[version=4]{mhchem}
\graphicspath{{figs/}{figsgaoerb/}} %

\usepackage{color}
\usepackage{epstopdf}
\usepackage{mathrsfs}
\DeclareMathAlphabet{\mathcal}{OMS}{cmsy}{m}{n}
\usepackage[citecolor=blue,linkcolor=blue,colorlinks=true]{hyperref}
\usepackage[utf8]{inputenc}
\usepackage[T1]{fontenc}

\usepackage{xcolor} 

\begin{document}%

\title{Predicting oscillations in complex networks with delayed feedback}

     \author{Shijie Liu$^{1,2}$, Jinliang Han$^{1}$, Jianming Liu$^{1}$, Tim Rogers$^{3}$}\email{t.c.rogers@bath.ac.uk}
     \author{Yongzheng Sun$^{1,2}$}\email{ yzsung@gmail.com}
\affiliation{
		\vskip 2mm
$^{1}$ School of Mathematics, China University of Mining and Technology, Xuzhou 221116,  China\\
$^{2}$ Shenzhen Research Institute, China University of Mining and Technology, Shenzhen 518057, China\\
$3$ Department of Mathematical Sciences, University of Bath, Bath BA27AY, UK}

\begin{abstract}
{\footnotesize \noindent} Oscillatory dynamics are common features of complex networks, often playing essential roles in regulating function. Across scales from gene regulatory networks to ecosystems, delayed feedback mechanisms are key drivers of system-scale oscillations. The analysis and prediction of such dynamics are highly challenging, however, due to the combination of high-dimensionality, non-linearity and delay. Here, we systematically investigate how structural complexity and delayed feedback jointly induce oscillatory dynamics in complex systems, and introduce an analytic framework comprising theoretical dimension reduction and data-driven prediction. We reveal that oscillations emerge from the interplay of structural complexity and delay, with reduced models uncovering their critical thresholds and showing that greater connectivity lowers the delay required for their onset. Our theory is empirically tested in an experiment on a programmable electronic circuit, where oscillations are observed once structural complexity and feedback delay exceeded the critical thresholds predicted by our theory. Finally, we deploy a reservoir computing pipeline to accurately predict the onset of oscillations directly from timeseries data. Our findings deepen understanding of oscillatory regulation and offer new avenues for predicting dynamics in complex networks.
\end{abstract}

	
\maketitle
\section{I. Introduction}
\noindent Oscillatory behaviors are widely observed across complex networked systems and often emerge near transitions characterized by reduced stability~\cite{may1972will,may2019stability,yang2023reactivity,baron2020dispersal,jiang2018predicting,escalante2025stability}. Such oscillations arise in diverse contexts, including neural activity, power-grid dynamics, social and technological networks, and ecological systems, where predator–prey interactions and population cycles provide a paradigmatic example~\cite{may1972limit,allesina2012stability,chowdhury2023eco,scheuerl2019asexual}. External variability—whether periodic forcing, stochastic perturbations, or slow environmental modulation—can trigger or reshape oscillatory dynamics, sometimes leading to abrupt or irreversible transitions~\cite{haurwitz1973diurnal,jia2021enhanced,irons2008fluctuations,2013Interactions,bathiany2018abrupt}. Oscillations may also be triggered by intrinsic stochasticity, particularly when close to a transition point \cite{krumbeck2021fluctuation,dai2012generic,scheffer2009early,dakos2015resilience,o2023early}. In this way, oscillatory dynamics may serve as early warning signals of declining resilience and impending functional loss~\cite{ceulemans2019effects}. 
Understanding and predicting the emergence of oscillatory behavior in complex networks is thus essential for anticipating transitions and preventing large-scale functional degradation.

A central mechanism shaping dynamics in many networked systems is memory, whereby past states influence present evolution. In ecosystems, this phenomenon is often termed \emph{ecological memory}, reflecting delayed demographic, environmental, or biogeochemical feedbacks~\cite{de2021dynamic,hendry1995role}. For example, soil organic matter decomposition depends not only on current conditions but also on past microbial and environmental states~\cite{ogle2015quantifying}, while tree growth reflects cumulative climatic influences, leading to the so-called prior growth effect~\cite{carrer2004age}. Analogous memory effects arise in neural adaptation, delayed control in engineering systems, and epidemic propagation in social networks~\cite{rogers2015assessing,moore2020predicting}. Mathematically, such memory can be incorporated through time delays or hereditary terms~\cite{anand2010ecological,golinski2008effects,kumar2025role,qin2025reordered}, which capture delayed responses and can induce oscillations~\cite{hastings2018transient,kuang1993delay,yang2023time,jiang2020hopf}. Yang et al. showed that short delays may enhance stability whereas long delays destabilize communities~\cite{yang2023time}, and Jiang et al. demonstrated Hopf bifurcation in delayed two-species systems beyond critical thresholds~\cite{jiang2020hopf}. Although low-dimensional systems have provided valuable insight into memory-driven oscillations, the interplay between memory effects and structural complexity in high-dimensional nonlinear networks remains largely unexplored~\cite{ma2023generalized}. 

Complex networks composed of many interacting units exhibit structural richness that plays a decisive role in shaping their collective dynamics. Increased structural complexity may undermine stability and drive systems toward instability~\cite{may1972will,may2019stability,yang2023reactivity,baron2020dispersal,jiang2018predicting,emary2022stability,feng2025complex} and drive other emergent phenomena including oscillations ~\cite{hu2022emergent,coclite2025replicator,bascompte2023resilience,yletyinen2016regime}. Ecological networks provide a well-studied realization of these principles, yet similar mechanisms operate in engineered, biological, and socio-technical networks. Modeling and predicting such behaviors remain challenging: large network size renders the system intrinsically high-dimensional, while heterogeneous interactions limit the applicability of traditional tools developed for low-dimensional or weakly nonlinear systems~\cite{ghadami2022data,poley2025interaction}. These challenges motivate the development of predictive frameworks capable of capturing high-dimensional, nonlinear dynamics in general complex networks ~\cite{gao2016universal,g2mp-bkhf}.

While theoretical models yield mechanistic insight, they often rely on assumptions—such as homogeneous interactions or simplified functional forms—that may not hold in real-world networks~\cite{gao2016universal}. This raises the broader question of how to predict complex network dynamics without imposing restrictive structural hypotheses. As a complementary strategy, data-driven machine learning approaches offer flexible, model-free frameworks capable of learning high-dimensional nonlinear dynamics directly from observations. In particular, reservoir computing has demonstrated strong performance in forecasting complex dynamical systems~\cite{xiao2021predicting,roy2022model,rakshit2023predicting,lee2023modelling,liu2025learning,yadav2025predicting}. Xiao et al. introduced parameter-aware reservoir computing to predict dynamics under parameter drift~\cite{xiao2021predicting}, and Roy et al. showed that echo state networks trained on a single attractor can reconstruct coexisting attractors at new parameter values~\cite{roy2022model}. 

\begin{figure*}[ht]
\begin{center}
\includegraphics[width=0.9\textwidth]{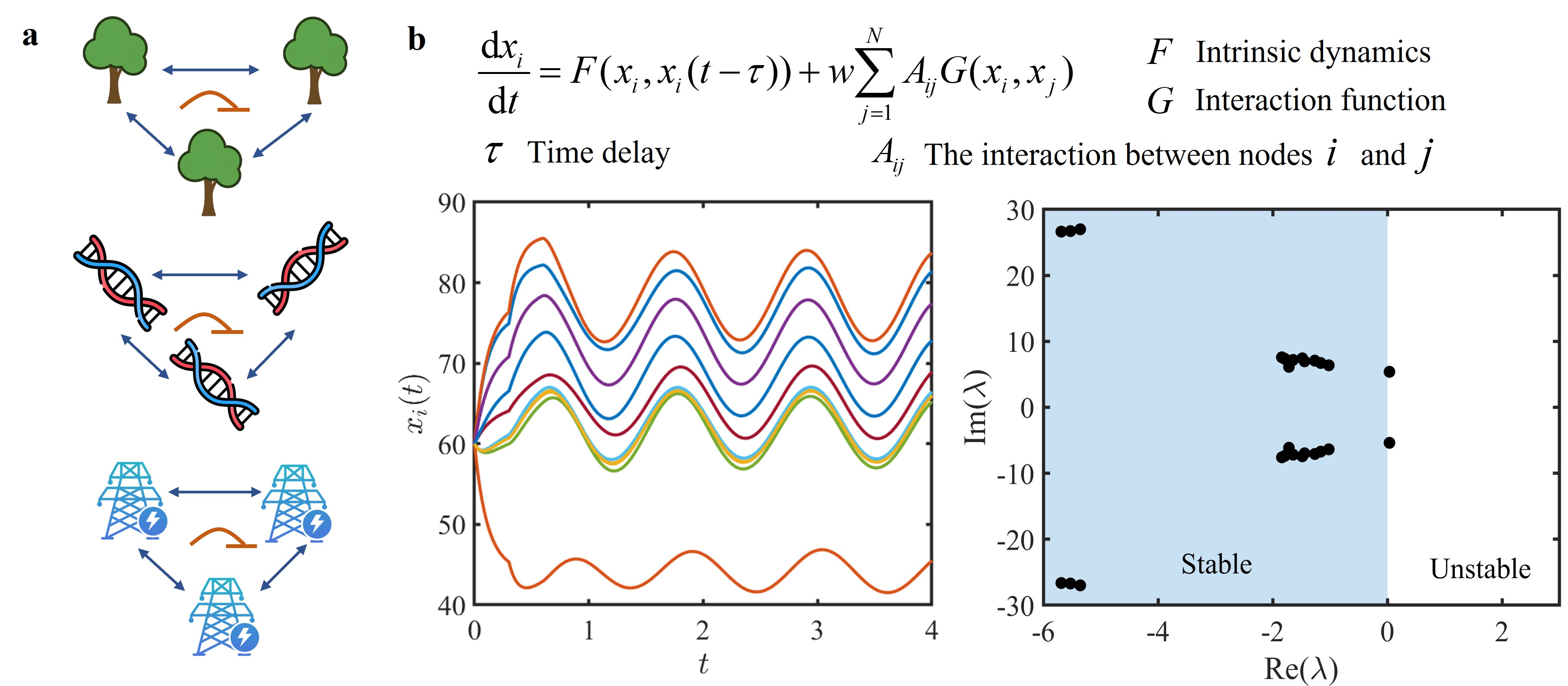}
\caption{{\bf Complex oscillations in dynamical systems.}
{\bf a.} Schematic illustration of real-world systems exhibiting oscillatory behavior: forest ecosystems (trees),  gene regulatory networks, and power grids. Blue arrows denote interspecific interactions, while orange lines represent intraspecific time-delay effects.
{\bf b.} General formulation of a complex time-delayed system. \emph{Left:} Temporal evolution of species abundances in system~(\ref{e1}), exhibiting sustained oscillatory dynamics. \emph{Right:} Linear stability analysis of system~(\ref{e1}). The system is stable when all eigenvalues of the Jacobian matrix have negative real parts, i.e., when they lie entirely in the left half of the complex plane. Eigenvalues associated with species are marked by black dots. The time delay destabilizes the system, as indicated by the emergence of purely imaginary eigenvalues. The underlying structure follows an ER network with $N=10$, connectivity $C=0.5$, and parameters $r_i=0.4$, $s_i=0.08$, and $\tau=0.3$.}
\label{fig1}
\end{center}
\end{figure*}

Here, we systematically investigate how structural complexity and memory jointly generate oscillatory dynamics in general complex networks, using ecological mutualistic networks as a representative example. Memory is incorporated as delayed intraspecific competition, and structural complexity is encoded in a random interaction network. We employ a dimension-reduction approach that maps the high-dimensional network dynamics onto an effective one-dimensional description while preserving essential stability properties~\cite{gao2016universal}. Our analysis reveals that the interplay between connectivity and delay induces oscillations via Hopf bifurcation, with stronger connectivity lowering the critical delay threshold. We go on to validate the findings of our theoretical framework through application to another setting -- an experimental study of an electronic circuit that reproduces oscillatory transitions beyond predicted thresholds. Finally, to address parameter uncertainty and structural heterogeneity, we apply reservoir computing to forecast oscillations driven by complexity and memory. The data-driven framework accurately predicts critical transitions, complements theoretical analysis, and generalizes to other classes of nonlinear networked systems, demonstrating a unified approach to oscillatory dynamics in complex networks.

\section{II. Modeling and prediction of oscillations}
\label{II}
\noindent In this section, we present the modeling framework and prediction approach for oscillatory dynamics. Figure~\ref{fig1}(a) shows schematic examples of oscillatory behavior in three representative real-world systems: forest ecosystems, gene regulatory networks, and power grids. To characterize memory effects on complex network dynamics, we construct a delayed network model in which memory length is represented by a time delay. Figure~\ref{fig1}(b) indicates a general formulation of such systems, where the first term describes intrinsic dynamics and the second term represents a general interaction function $G$. $\omega$ is the interaction strength.

In the development of our theory, we will focus primarily on the application to ecological networks, in which dynamical variables $x_i(t)$ denote the abundance of species $i$ at time $t$. Mathematically, our model is formulated as follows:
\begin{equation}\label{e1}
\frac{\mathrm{d}x_i(t)}{\mathrm{d}t}=x_i(t)\left(r_i- s_i x_i(t-\tau)\right)+
\frac{\sum_{j=1}^{N}a_{ij}x_{i}(t)x_j(t)}{D_i+E_i x_i(t)},
\end{equation}
where the time delay $\tau > 0$ represents the length of ecological memory. The parameter $r_i$ is the intrinsic growth rate of species $i$, and $s_i$ controls the strength of intraspecific competition for resources, which depends on the delayed density $x_i(t - \tau)$. The second term reflects saturated mutualistic interactions. Specifically, $a_{ij} = 1$ for $i \neq j$, indicating mutualism between species $i$ and $j$, and $a_{ij} = 0$ otherwise. The parameters $D_i>0$ and $E_i>0$ determine the saturation level of these mutualistic interactions~\cite{gao2016universal}. The initial condition of model (\ref{e1}) is $x_i(t)=\varphi(t)\geq 0$,  $t\in[-\tau,0]$,
where $\varphi\in C([-\tau,0],[0,+\infty])$, and $\varphi(0)>0$. The model captures potential delayed effects, enabling a systematic analysis of how ecological memory influences the network dynamics of the system~\cite{anand2010ecological,golinski2008effects,ogle2015quantifying}.

To explore how ecological memory and network topology jointly shape oscillatory dynamics, we couple species through Erd\H{o}s-R\'{e}nyi (ER) random networks~\cite{erd6s1960evolution}. The time series of system~(\ref{e1}) are shown in Fig.~\ref{fig1}(b). For complex systems, stability can be evaluated by the largest real part of the eigenvalues of the Jacobian matrix $\mathcal{J}$, denoted $\mathrm{Re}(\lambda_1)$. The system is stable when $\mathrm{Re}(\lambda_1) < 0$, as illustrated in Fig.~\ref{fig1}(b). Then, we further investigate the oscillatory behavior of the system. Figure~\ref{fig2}(a) presents the dynamic evolution of system~(\ref{e1}) over time. The results reveal that increasing connectivity and delay jointly destabilize the system, inducing oscillatory behavior and underscoring their combined influence on high-dimensional, nonlinear dynamics. Moreover, for a fixed network structure, a critical delay exists that triggers oscillations, raising the question of how this threshold can be predicted.

\subsection{Derivation of theoretical critical delay}
In the following, we derive the critical delay that induces oscillations. Since the complexity of high-dimensional and nonlinear dynamics in the model, we reduce the system to a lower-dimensional form using the Gao-Barzel-Barab\'{a}si (GBB) framework ~\cite{gao2016universal}. This dimension reduction preserves the key dynamical and critical properties of the original system, enabling a more tractable theoretical analysis and facilitating the identification of the critical delay at which oscillations emerge~\cite{ma2023generalized,gao2016universal}. The corresponding reduction operator is defined as follows: 
$\mathcal{L}(x)=\sum_{i,j}a_{ij}x_j/\sum_{i,j}a_{ij}$
The operator $\mathcal{L}$ takes the vector $x$ as input and outputs a weighted average of its nearest-neighbor components, with weights proportional to the degree of each component. By using the operator, we obtain the following effective one-dimensional (1D) system
\begin{equation}\label{e2}
\frac{\mathrm{d}x_{\mathrm{eff}}(t)}{\mathrm{d}t}=x_{\mathrm{eff}}(t)\left(r-s x_{\mathrm{eff}}(t-\tau)
+\frac{\beta_{\mathrm{eff}}x_{\mathrm{eff}}(t)}{D+Ex_{\mathrm{eff}}(t)}\right),
\end{equation}
where $x_{\mathrm{eff}}(t)$ is the effective abundance, and the role of the network is compressed into a single effective parameter 
$\beta_{\text{eff}}=\sum_{i,j,k}a_{ij}a_{kj}/\sum_{i,j}a_{ij}$. 
This reduction collapses $N^2$ microscopic parameters $a_{ij}$ into a single macroscopic parameter $\beta_{\mathrm{eff}}$, while preserving the network's structural features through degree-weighted average. We further compare the dimensionality reduction method with other approaches, and the results are presented in the Methods section.

The model has two equilibria, $x_{\mathrm{eff}}=0$ and $x_{\mathrm{eff}}^{*}=(a_1+\sqrt{a_1^2+4rsDE})/2sE$, where $a_1=rE+\beta_{\mathrm{eff}} - sD$.
Next, we conduct a theoretical analysis and focus on the positive equilibrium $x_{\mathrm{eff}}^{*}$. The oscillatory behavior of delayed systems can usually be analyzed through their characteristic equations. Specifically, the characteristic equation of Eq.~(\ref{e2}) at $x_{\mathrm{eff}}^{*}$ is given by
\begin{equation}\label{e3}
\lambda+B_1 e^{-\lambda \tau}-B_2=0,
\end{equation}
where $B_1=s x_{\mathrm{eff}}^{*}$, $B_2=(\beta_{\mathrm{eff}} Dx^{*}_{\mathrm{eff}})/(D+Ex^{*}_{\mathrm{eff}})^{2}$.
According to the equation, we can examine the Hopf bifurcation in system~(\ref{e2}) to derive the critical delay theoretically. Let $\lambda=i\omega$, the critical delay can then be obtained by  analysis as follows (see Methods section):
\begin{equation}\label{e4}
\begin{split}
\tau^{*}=\arccos\Big(\frac{\beta_{\mathrm{eff}} Dx^{*}_{\mathrm{eff}}}{sx_{\mathrm{eff}}^{*}(D+Ex^{*}_{\mathrm{eff}})^{2}}
\Big)/\omega.
\end{split}
\end{equation}

In particular, for undirected networks with node degrees $k_i$ that are uncorrelated, we have $\beta_{\rm{eff}}=\langle k^2 \rangle/\langle k \rangle$. For random ER networks, $\langle k^2 \rangle = \langle k \rangle (\langle k \rangle + 1)$ \ yielding  $\beta_{\rm{eff}} = \langle k \rangle + 1 = N C + 1$. We use Fig.~\ref{fig2}(b) to show the states of the original system~(\ref{e1}) along with the analytically derived expressions~(\ref{e4}) across different values of $\tau$ and $C$. By comparison, it can be seen that the theoretical results are in good agreement with the numerical simulations, indicating that the theoretical analysis can effectively predict the oscillatory behavior of the high-dimensional system. 

\subsection{Oscillations in synthetic and real networks}
\begin{figure*}[ht]
\begin{center}
\includegraphics[width=0.9\textwidth]{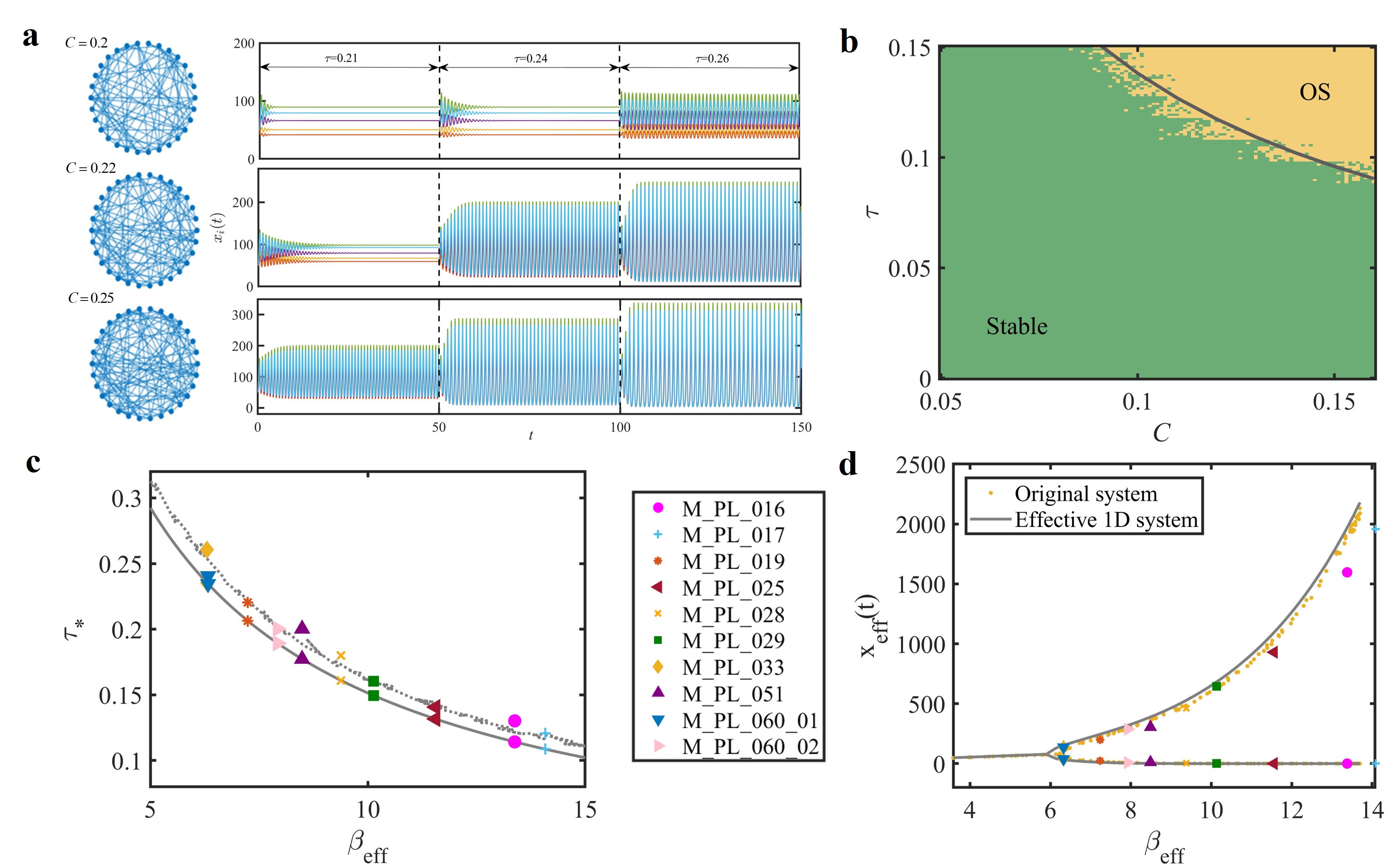}
\caption{ {\bf Critical delay for oscillations induced by network complexity and time delay.} {\bf a.} The time series of system~(\ref{e1}) under various connectivity and delays.  For $C = 0.2$, the species abundance remains stable at $\tau = 0.21$ and $\tau = 0.24$, but exhibits oscillatory behavior at $\tau = 0.26$.  As connectivity increases to $C = 0.22$, the system also transitions from a stable to an oscillatory regime, with oscillations already appearing at $\tau = 0.24$. When $C = 0.25$, oscillations emerge as early as $\tau = 0.21$, and their amplitude further increases with larger delays. {\bf b.} The colored regions correspond to the dynamical states of the original high-dimensional system~(\ref{e1}) on ER random networks, whereas the solid lines depict the theoretical estimates~(\ref{e4}) derived from the reduced one-dimensional system~(\ref{e2}). As $\tau$ and $C$ increase, the system states shift from stability to oscillatory behavior, as shown by the green (stable) and yellow (oscillatory, OS) regions. {\bf c.} The reduced one-dimensional model~(\ref{e2}) predicts the critical delay of the original high-dimensional system~(\ref{e1}). Increasing $\beta_{\mathrm{eff}}$ decreases the critical delay of the system. Both theoretical (solid lines) and numerical (dotted lines) results are shown for ER networks with varying connectivity $C$ and network size $N = 100$, while different markers represent results from ten empirical networks displayed around the panel. {\bf d.} The reduced 1D model accurately reproduces the abundance dynamics of the original high-dimensional model for both ER networks and empirical mutualistic networks. The network structures are the same as those used in the left panel. In panel~(a), $N = 100$; for panels~(b) and~(c), $r_i=0.4$, $s_i=0.08$, $D_i=1$, $E_i=1$, and the networks are ER random networks.}
\label{fig2}
\end{center}
\end{figure*}
Further, we verify the validity of the theoretical critical time delay. The critical delay as a function of $\beta_{\mathrm{eff}}$ is shown in Fig.~\ref{fig2}(c), where the solid line represents the theoretical prediction for ER random networks derived from Eq.~(\ref{e2}), while the dotted line shows the corresponding numerical results from Eq.~(\ref{e1}). The two critical delays are basically consistent. This indicates that the one-dimensional simplified system (\ref{e2}) can be used to predict the oscillations of the original high-dimensional system (\ref{e1}). When the reduced system undergoes oscillations, the high-dimensional system correspondingly exhibits oscillatory behavior. Moreover, we examine the critical delay in the system (\ref{e1}) with empirical mutualistic networks~\cite{jiang2018predicting}, and compare it with that of the reduced system (\ref{e2}). The results remain consistent, illustrating the robustness of the reduced model in capturing the onset of oscillations. To assess the accuracy of the reduction more directly, Fig.~\ref{fig2}(d) compares the state evolution between system~(\ref{e1}) and system~(\ref{e2}). The effective variable $x_{\mathrm{eff}}(t)$ of the high-dimensional system~(\ref{e1})
is constructed as a weighted average of the node states $x_i(t)$. Under the same parameter settings, both systems exhibit nearly identical state evolution, with minor differences in the location of the critical points.
\begin{figure*}[ht]
\begin{center}
\includegraphics[width=1\textwidth]{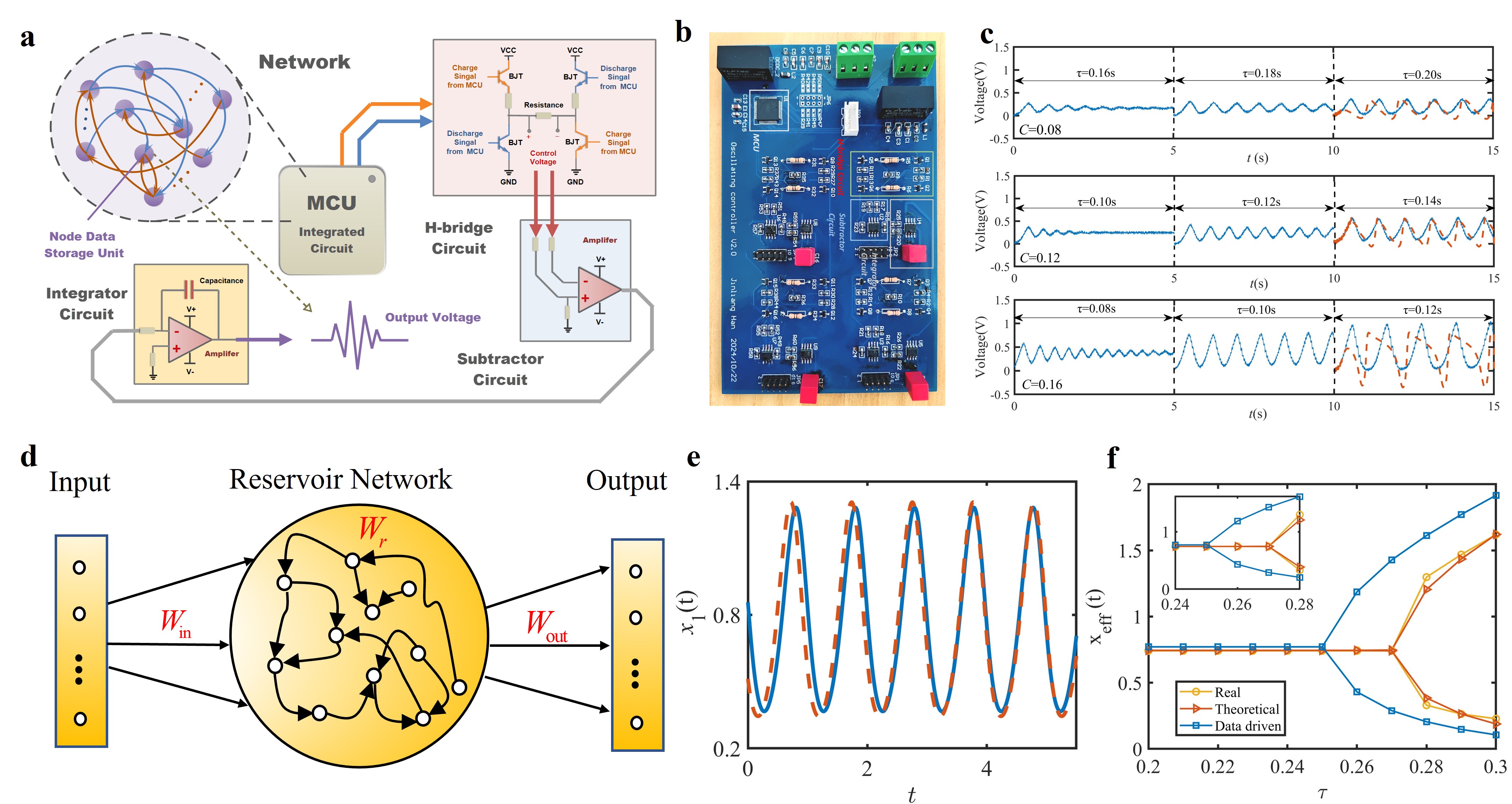}
\caption{\textbf{Prediction and validation of oscillatory behavior.} 
{\bf a.} The design block diagram of the oscillation control circuit experimental system consists of the MCU chip, the H-bridge driver circuit, the subtractor circuit, and the integrator circuit connected in series. The network nodes and their topological information are stored in the registers of the MCU. {\bf b.} The physical PCB circuit, realized based on the design block diagram, is capable of simultaneously detecting the state waveforms mapped from up to four nodes. {\bf c.} Experimental voltage u for various delays and ER random networks with different connectivity: $C = 0.08$, $C = 0.12$, and $C = 0.16$. The dashed line represents the predictions obtained from reservoir computing. All networks have $N = 100$ nodes, and model parameters are set to $r_i = 0.4$, $s_i = 0.08$, and $D_i = E_i = 1$. {\bf d.} Schematic diagram of the reservoir computing architecture, which includes three layers: input, reservoir (hidden), and output. The reservoir is trained using time series corresponding to three delays ($\tau = 0.23$, $0.24$, and $0.25$) and demonstrates generalization beyond the training range by adjusting the input delay parameter. {\bf e.} Prediction of oscillatory dynamics in system~(\ref{e1}).  Time series of $x_1(t)$ from system~(\ref{e1}) and its prediction at $\tau = 0.28$ are shown to validate the accuracy and robustness of the training. {\bf f.} Effective system abundance $x_{\mathrm{eff}}(t)$ as a function of $\tau$ for three cases. Theoretical results are obtained using the GBB method (system~(\ref{e4})), real data are generated from system~(\ref{e1}), and data-driven results are produced by the reservoir computing model.
}\label{fig3}
\end{center}
\end{figure*}

\section{III. Experimental test and Data-Driven Prediction of Oscillations}

\subsection{Experimental realization using programmable micro-controller} 
To experimentally validate our predictions, we employ an experimental system designed around circuit simulation~\cite{zhang2025non}. The structural composition of the network oscillation control circuit system is formed by successively connecting in series a microcontroller unit (MCU) chip, an H-bridge drive circuit, a subtractor circuit, and an integrator circuit, as depicted in Fig.~\ref{fig3}(a). Correspondingly, a printed circuit board (PCB) has been fabricated, as shown in Fig.~\ref{fig3}(b). The information regarding network nodes and their topological structures is encoded and stored within the registers of the MCU. Simultaneously, by leveraging its internal register set, a delay controller is constructed. This enables precise regulation of time delays and dynamic storage of node state information. The MCU generates level-conversion instructions in real time according to the right-hand-side function of Eq.~($\ref{e1}$). These instructions are used to perform closed-loop regulation of the physical quantities of the peripheral circuits (details can be found in the Methods section). This ensures that the evolutionary process of the circuit state can accurately represent the dynamic behavior of the node state. Consequently, the output of the system can visually exhibit the dynamic characteristics and control mechanisms of complex networks with delay. This significantly enhances the verifiability of theoretical models and provides stronger experimental support. 

We randomly generated three types of network structures with different connection probabilities $p$, $N = 100$. By adjusting the system delay within the MCU, we monitored the dynamic response of the circuit output voltage in real-time. This was done to determine whether oscillations occurred in the network and to analyze the characteristics of its amplitude regulation. Taking the second node as the object of observation, we used an oscilloscope to collect and record signals. The results are presented in Fig.~\ref{fig3}(c). When $C = 0.08$, at $\tau=0.16$, the voltage signal tended towards a static equilibrium state. However, when $\tau = 0.18$, distinct oscillations emerged, indicating that the critical delay time $\tau^{*}$ for the onset of oscillations satisfies $0.16<\tau^{*}<0.18$. Substituting the relevant network parameters into formula ($\ref{e1}$), the critical delay time of the reduced system was approximately $\tau^{*}=0.18$. Analogously, for $C = 0.12$, the critical delay range was $0.1<\tau^{*}<0.12$, corresponding to the theoretically predicted value of $\tau^{*} = 0.12$. When $C = 0.16$, the range $0.08 < \tau<0.1$ was observed. Combining with formula ($\ref{e2}$), the critical delay time of the reduced system was $\tau^{*}=0.09$. These results demonstrate that the oscillation behavior of
high-dimensional networks can be effectively predicted by the critical delay time of a one-dimensional simplified model. Moreover, it was further discovered that as the delay increases, the oscillation amplitude gradually enlarges, and this effect is more pronounced at higher connection densities. This finding is highly consistent with the simulation results shown in Fig.~\ref{fig2}. Through the construction of a circuit system capable of mapping the dynamics of complex networks, this experiment successfully reproduced the dynamic process from amplitude death to the restoration of oscillations. This not only validates the possible oscillation mechanisms in complex networks but also provides a means for rapidly predicting their future fluctuation trends (see the supplementary material video).

\subsection{Reservoir computing pipeline}
\noindent Although previous approaches can predict system oscillations, their practical application is often constrained by the difficulty of parameter estimation. To overcome this limitation, we adopt a numerical framework based on reservoir computing, which learns the system's temporal dynamics and predicts oscillatory behavior without requiring explicit knowledge of model parameters.

As a powerful data-driven approach, reservoir computing effectively captures the nonlinear characteristics of complex systems and reproduces their dynamical properties, providing a practical alternative to traditional methods by avoiding intricate theoretical derivations~\cite{roy2022model,rakshit2023predicting,liu2025learning}. The scheme of reservoir computing is shown in Fig.~\ref{fig3}(d), where the dynamics are described with the following map:
\begin{equation}\label{e5}
r(t+1)=(1-\alpha)r(t)+\alpha\tanh[W_{\mathrm{r}}\cdot r(t)+W_{\mathrm{in}}\cdot u(t) ],
\end{equation}
where $r(t)=[r_1(t),r_2(t),\cdots,r_{N}(t)]$ represents the state of the reservoir at time $t$. $\alpha\in (0,1)$ is the leakage rate. $W_{\mathrm{r}}\in R^{ N\times N}$ is a random sparse matrix, generated by an ER graph configuration. An $N$-dimension signal inputs through an $N\times M$ input-weighted matrix $W_{\mathrm{in}}$. $u(t)\in R^{M+1}$ is an $M+1$ dimensional input vector, where the first $M$ elements are the dynamics of the input signal $\tilde{u}(t)$ at time $t$ and the last element is the corresponding parameter, i.e., $u(t)=[\tilde{u}(t) p]^{T}$. We randomly choose the elements of $W_{\mathrm{in}}$ from the uniformly distributed interval $[-\sigma,\sigma]$ and form $W_{\mathrm{in}}$ in such a way that the information of the input data would be equally distributed to the nodes of the reservoir. Each input dimension is projected onto $N/(M-1)$ reservoir nodes, while parameter information is embedded across all nodes. This design allows the reservoir to capture the interplay between system dynamics and their associated parameters.

We train the machine for $N_p$ different parameter values one by one and successively store the values of $r(t)$, and the initial state is set to be $r(0)=0$. It is conventional to define a new vector $\tilde{r}$ with the same odd row elements as those of $r$ but with the even row elements as the squared values of the corresponding even row elements of $r(t)$. We store the values of $\tilde{r}(t)$ within the matrix $R$ after ignoring $t_{b}$ time points as the transient dynamics of the reservoir. Therefore, $R$ is a matrix with dimension $N \times N_p(t_f-t_{b})$ for training data of length $t_f$. After calculating $R$, we can get the readout weight matrix from the system of a linear equation, $U=W_{\mathrm{out}}R$,
where $U$ is a matrix having targeted data sets with dimension $M\times N_p(t_f-t_{b})$. Here, the target output corresponds to the system state at the next time step. To determine the readout matrix $W_{\mathrm{out}}$, we use Ridge regression to minimize the discrepancy between $U$ and $W_{\mathrm{out}}R$:
\begin{equation}\label{e6}
W_{\mathrm{out}}=U R^{T}(RR^{T}+\beta I)^{-1},
\end{equation}
where $\beta$ is the regularization term that prevents overfitting. Once training is completed, the model can generate predictions solely based on the parameter values.

During the prediction phase, the input $u(t)$ is replaced by the output $v(t)$, creating a closed-loop system that allows the reservoir to evolve autonomously. The system state is then updated iteratively from $v(t)$ to $v(t+dt)$ as follows:
\begin{equation}\label{e7}
\begin{split}
r(t+1)&=(1-\alpha)r(t)+\alpha\tanh[W_{\mathrm{r}}\cdot r(t)+W_{\mathrm{in}}\cdot\nu(t)],\\
\nu(t)&=\left[\hat{\nu}(t) p_{\mathrm{new}}\right]^{T},\\
\hat{\nu}(t)&=W_{\mathrm{out}}\tilde{r}(t).
\end{split}
\end{equation}
By iterating Eq.~(\ref{e7}), the reservoir generates predicted time series for new parameter values. The prediction process begins from an initial condition corresponding to a bounded solution, often chosen from the training trajectory. Prior to generating predictions, the reservoir undergoes a warm-up phase to eliminate the influence of its initial state. The scheme involves several hyperparameters-spectral radius, leaking rate, input weight range, and regularization parameter-which are optimized using a simultaneous optimization technique based on the root-mean-square error (RMSE) of the predicted series relative to the target, averaged over multiple reservoir realizations. Once the optimal hyperparameters are obtained, the final reservoir is constructed, ready to predict system dynamics under new parameter values and initial conditions.

\subsection{Proof of concept test}
Then, we use this method to predict the oscillation of the system (\ref{e1}) for ER random networks with $N=100$. We employ the Runge-Kutta method to generate data at $\tau=0.23,0.24,0.25$  by system (\ref{e1}) with a step size of $\mathrm{d}t=0.011$ for model training. Under these time-delay parameters, the system exhibits a stable state. The number of training data at each parameter value $\tau$  is $10000$. The parameter settings for the model are as follows: the size of the reservoir is $1000$, the sparsity of the reservoir weight matrix is $0.1$, the leakage rate is $0.3$, and the regularization parameter of Ridge regression is $10^{-5}$. The reservoir weight matrix is first randomly generated through uniform distribution in the interval $[-0.5,0.5]$, then regularized with a spectral radius of $0.8$.

We investigate the system behavior at $\tau=0.26$ and $\tau=0.28$. The results show that the system remains in a stable state at $\tau=0.26$, which is accurately captured by the model. When $\tau=0.28$, the system enters an oscillatory regime, and the predicted results agree well with the actual dynamics. The predicted results of $x_1(t)$ in the system (\ref{e1}) at $0.28$ are shown in Fig.~\ref{fig3}(e). To further predict the critical delay for the system to transition from a stable state to an oscillatory state, we conduct predictions at $\tau=0.26,0.261,0.262,...,0.278,0.279$  and $0.28$. The results show that the system is in a stable state at $\tau=0.274$  and an oscillatory state at $\tau=0.275$, thus the predicted critical delay of oscillation is $0.275$. We see that the model results are consistent with the predicted results, indicating that the prediction method based on reservoir computing is effective. Applying this method to predict the dynamics of the circuit system yields the results shown in Fig. \ref{fig3}(c). We can observe that the reservoir computing approach is also capable of accurately predicting the dynamics originating from the circuit.

As shown in Fig.~\ref{fig3}(f), we further compared the trajectories of $x_{\mathrm{eff}}(t)$ obtained by different methods, using the GBB framework as a theoretical reference. The results indicate that the one-dimensional theoretical model detects the onset of oscillations earlier, providing an effective early-warning signal for high-dimensional systems, whereas reservoir computing delivers highly accurate predictions of the actual critical delay. Together, these findings highlight the complementary nature of the two approaches: dimension reduction offers theoretical guidance, while reservoir computing provides data-driven predictions, and their combination enhances the overall reliability of oscillation forecasting.

\section{IV. Discussion}
\noindent This study investigates the memory effect in intraspecific interactions within complex network models, with a focus on how structural complexity and time delays influence the emergence and characteristics of oscillatory dynamics~\cite{yang2023time}. To address the challenges of predicting oscillations in high-dimensional ecological networks, we first applied dimension reduction to transform the system into a one-dimensional representation, enabling analytical tractability. Theoretical analysis of this reduced system allows us to derive an explicit expression for the critical delay, which serves as a threshold for the onset of oscillatory behavior. Using this critical delay, we successfully predicted oscillations in the original high-dimensional networks and demonstrated that, as network connectivity increases, the critical delay decreases, leading to earlier oscillation onset. Comparisons with other dimension-reduction approaches indicate that the method presented here is the most effective. In addition, we employed reservoir computing as a data-driven framework to predict oscillatory dynamics in complex high-dimensional systems, achieving accurate forecasts without requiring explicit knowledge of all model parameters. Together, these results provide a versatile framework for analyzing and predicting oscillations in complex ecological networks, highlighting the combined power of theoretical and machine-learning approaches.

Despite the success of the dimension reduction framework in predicting system oscillations, its generalizability across diverse dynamical systems remains to be explored. Extending this approach to a broader range of network structures and interaction types will be essential for establishing its robustness and wider applicability. Similarly, while reservoir computing has demonstrated strong predictive capability for systems based on empirical mutualistic networks, its current implementation relies primarily on training data generated from theoretical models. Future studies should incorporate empirical observations to validate the practical performance of the framework~\cite{arani2021exit,hu2022emergent}. Bridging the gap between theoretical models and real-world data will enhance the reliability of oscillation predictions and provide valuable insights for understanding dynamics in ecological and other complex networked systems.

\section*{Acknowledgements}
\noindent This work was supported by the National Natural Science Foundation of China (NSFC) (Grant No.~12271519), the Fundamental Research Funds for the Central Universities (Grant Nos.~2023ZDYQ11005 and 2024KYJD2002).

\section*{Competing interests}
\noindent The authors declare that they have no competing interests.

\section*{Data availability}
\noindent 
All real ecological network data supporting the findings of this study are publicly available from the Web of Life databaseat https://www.web-of-life.es/. The simulation code supporting this work is available for download at https://github.com/Lsj409/delay-systems/tree/main.

\section*{Appendix: Methods}
\subsection{1. Derivation of the critical delay}
\noindent Here, we present the detailed derivation of Eq.~(\ref{e4}). Assuming $\lambda=i\omega$, we have
\begin{align*}
i\omega+B_1 e^{-\lambda \tau}-B_2=0,
\end{align*}
Then, the real and imaginary parts of the equation are
\begin{equation}
\begin{split}\label{e8}
B_1\cos\omega\tau=B_2,\
B_1\sin\omega\tau=\omega.
\end{split}
\end{equation}
Hence, we obtain
\begin{align*}
\omega^2=B_1^2-B_2^2.
\end{align*}
We set $\omega>0$, then for $B_2>0$,
\begin{align*}
\tau^{k}_{i}=\frac{\arccos\frac {B_2}{B_1}+2k\pi}{\omega}=
\frac{\arcsin\frac{\omega}{B_1}+2k\pi}{\omega},
\ \ k=0,1,2,\cdots,
\end{align*}
and for $B_2<0$,
\begin{align*}
\tau^{k}_{i}=\frac{\arccos\frac{B_2}{B_1}+2k\pi}{\omega}=
\frac{\pi-\arcsin\frac{\omega}{B_1}+2k\pi}{\omega},
\ \ k=0,1,2,\cdots.
\end{align*}
Due to the periodic nature of inverse trigonometric functions, the critical values of time delay and their corresponding critical frequency values obtained through the method above also exhibit periodicity. Considering the practical significance of time delay in real life, we only take the first positive value of $\tau$ in the solution as the critical value of time delay. Then, we can obtain the critical delay as Eq.~(\ref{e4}).
Let $\lambda(\tau)=\xi(\tau)+i\omega(\tau)$ be the eigenvalue of Eq.~(\ref{e3}). Then, to find the transversality condition for the occurrence of Hopf-bifurcation, we need to verify $\frac{\mathrm{d}\xi(\tau)}{\mathrm{d}\tau}\Big|_{\tau=\tau^*}>0$. Taking $\lambda$ into the Eq.~(\ref{e3}), and equating real and imaginary parts are
\begin{equation}\label{e9}
\begin{split}
\xi+B_1\exp(-\xi\tau)\cos\omega \tau-B_2=0,\\
\omega-B_1\exp(-\xi\tau)\sin\omega\tau=0.
\end{split}
\end{equation}
Differentiate (\ref{e9}) and set $\xi=0$, we have
\begin{equation}\label{e10}
\begin{split}
X(\omega)\frac{\mathrm{d}\xi}{\mathrm{d}\tau}
+Y(\omega)\frac{\mathrm{d}\omega}{\mathrm{d}\tau}=Z(\omega),\\
-Y(\omega)\frac{\mathrm{d}\xi}{\mathrm{d}\tau}
+X(\omega)\frac{\mathrm{d}\omega}{\mathrm{d}\tau}=W(\omega),
\end{split}
\end{equation}
where $X(\omega)=1- B_1\tau \cos \omega\tau$, $Y(\omega)=B_1\tau \sin \omega\tau$, $Z(\omega)=B_1\omega \sin \omega\tau$, $W(\omega)=B_1\omega \cos \omega\tau$.
Through (\ref{e10}), we obtain
\begin{align*}
\frac{\mathrm{d}\xi(\tau)}{\mathrm{d}\tau}\Big|_{\tau=\tau^*}
=\frac{Z(\omega)X(\omega)-Y(\omega)W(\omega)}{X^2(\omega)+
Y^{2}(\omega)}.
\end{align*}
When $Z(\omega)X(\omega)-Y(\omega)W(\omega)>0$, the system will undergo Hopf-bifurcation.

\subsection{2. Dimension reduction methods}
\begin{figure*}
\begin{center}
\includegraphics[width=1\textwidth]{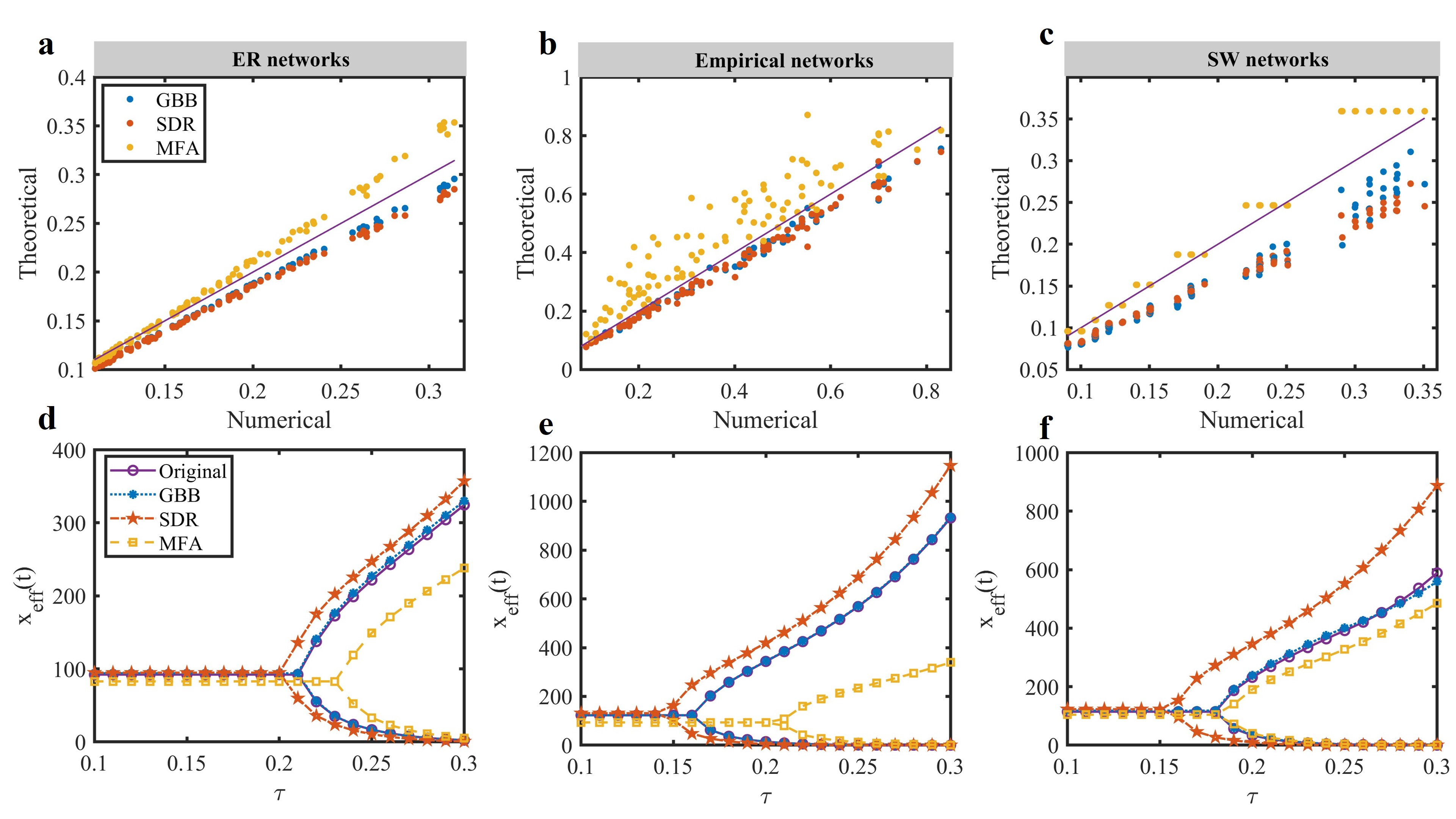}
\caption{ {\bf Comparison of three dimension reduction methods.}
{\bf a-c.} The critical delay $\tau^*$ of the original high-dimensional system~(\ref{e1}) is compared with that obtained from three reduced representations: the GBB system~(\ref{e2}), the SDR system~(\ref{e13}), and the MFA system~(\ref{e14}). Panel (a) shows results for ER random networks with $N=100$ and varying connectivity $C$, panel (b) for empirical networks, and panel (c) for SW networks with $N=100$, varying nearest neighbors $K$ and $C=0.15$. Numerical results are obtained from direct simulations, while theoretical results represent the predicted critical delays from Eq.~(\ref{e4}) for the reduced systems. The solid line with slope 1 serves as a reference for perfect agreement.
{\bf d-f.} Time series of the effective system abundance $x_{\mathrm{eff}}(t)$ are shown for the same four systems (the original system and the three reduced forms) under representative ER, empirical, and SW networks, with varying delay $\tau$, respectively. These panels highlight how the dynamics of each network type are reproduced across the different dimension reduction methods. Model parameters are fixed at $r_i = r=  0.4$, $s_i = s = 0.08$, $D_i = D = 1$, and $E_i = E = 1$.}\label{fig4}
\end{center}
\end{figure*}

To assess the prediction accuracy of dimension reduction techniques, we employ two alternative dimension reduction methods, the spectral dimension reduction (SDR)~\cite{laurence2019spectral} and the mean-field approximation (MFA)~\cite{nakao2010turing}, to determine the critical delay. The predictions from all three methods are compared in Figs.~\ref{fig4}(a)-(c), where closer proximity to the diagonal line indicates smaller errors. In Fig.~\ref{fig4}(a), the MFA systematically overestimates the critical delay relative to the original system, highlighting its limited accuracy in high-dimensional networks. By contrast, both GBB and the SDR method yield more accurate predictions, with the former showing minimal deviation, as it effectively preserves the structural properties of the network. Figures~\ref{fig4}(b) and (c) further confirm these trends across a variety of empirical mutualistic networks and small-world (SW) networks~\cite{watts1998collective}. To further evaluate their performance, we select one representative network from each class. The changes in system state with delay are shown for an ER random network, an empirical network, and an SW network in Figs.~\ref{fig4}(d)-(f), respectively. The SDR consistently underestimates the critical delay compared with the original system. The MFA overestimates the delay in both ER and empirical networks, although its prediction in the SW network is closer to that of the original system. In contrast, the GBB reduced method provides critical delays that closely match those of the original system across all three network types, thereby reaffirming the robustness and effectiveness of the GBB approach. Having established its superior performance, we next employ the GBB method to examine oscillation prediction under environmental perturbations.
\noindent The GBB method has been described in the main text, here, we introduce two additional dimension reduction approaches. For spectral dimension reduction, we consider $R(x)=\sum_{i=1}^{N}a_ix_i$, where $a_i$ is the element of the eigenvector corresponding to the maximum eigenvalue of the adjacency matrix $A$~\cite{laurence2019spectral}. This method captures the dominant network mode that governs the system's dynamics. Applying operator $R$ on both sides, we can write Eq.~(\ref{e1}) as
\begin{equation}\label{e11}
\begin{split}
& R\left(\frac{\mathrm{d}x_i(t)}{\mathrm{d}t}\right)=R\Big(x_i(t)\Big(r_i- s_i x_i(t-\tau) +\frac{\sum_{j=1}^{N}a_{ij}x_j(t)}{D_i+E_ix_i(t)}\Big)\Big),\\
&\approx R(x)(r_i-R(s_ix_i(t-\tau))+
\left(\frac{R(\sum_{j=1}^{N}a_{ij}x_i(t)x_j(t))}{R(D_i+E_i x_i(t))}\right).
\end{split}
\end{equation}
After simplification, we have
\begin{equation}
\begin{split}\label{e12}
\frac{\mathrm{d}R(x)}{\mathrm{d}t}=&R(x)(r_i-R(s_i x_i(t-\tau)))+
R(k)\frac{R(x)R(x)}{D_i+E_i R(x)}.
\end{split}
\end{equation}
By substituting $R(k)=\alpha_1$, $r_i=r$, $s_i=s$, $D_i=D$ and $E_i=E$ into Eq.~(\ref{e12}), then we can obtain a one-dimensional reduction of the original $N$-dimensional dynamical system as
\begin{equation}\label{e13}
\begin{split}
\frac{\mathrm{d}R(x)}{\mathrm{d}t}=&R(x)\Big(r- sR(x(t-\tau))+
\frac{\alpha_1 R(x)}{D+E R(x)}\Big).
\end{split}
\end{equation}

For mean-field approximation~\cite{iacopini2019simplicial}, the system (\ref{e1}) can be written as
\begin{equation}\label{e14}
\frac{\mathrm{d}x(t)}{\mathrm{d}t}=x(t)\left(r-s x(t-\tau)
+\frac{\langle k \rangle x(t)}{D+Ex(t)}\right),
\end{equation}
where $\langle k \rangle $ is the average degree of matrix $A$. It can be seen that the Eq.~(\ref{e13}) and Eq.~(\ref{e14}) is similar with Eq.~(\ref{e2}), but the specific networks parameters are different.

\subsection{3. Validation Methods for Electronic Circuit Experiments}
\noindent
In the circuit, the MCU precisely regulates the resistance voltage at the central node of the H-bridge drive circuit via external I/O pins. This voltage signal is first processed by a subtractor and then fed as an input into the integrator circuit. The dynamic evolution of the capacitor voltage $\Delta u$ within the integrator serves as an effective indicator of the oscillation behavior of network nodes. By adjusting the delay step $m$ in the MCU's delay register, a systematic investigation of the oscillation characteristics of network nodes under various time delays can be carried out.

The input terminals of the H-bridge circuit are categorized into two sets of control signals for charging and discharging. The operational state of the circuit is determined by the sign function $\mathrm{sgn}_{\Delta u_i}$, which is defined as follows:
\begin{equation}
\begin{split}\label{e15}
\mathrm{sgn}_{\Delta u_i} = \mathrm{sgn} \left( r_i x_i(n) \left( 1 - s_i x_i(n - m) \right)+\frac{\sum_{j = 1}^{N} a_{ij} x_i(n) x_j(n)}{D_i + E_i x_i(n)} \right).
\end{split}
\end{equation}
Here, $x_i(n)$ represents the state variable of node $i$ at the discrete time instant $n$. Given a time step of $h$, $x_i(n - m)$ corresponds to the historical state of node $i$ at a delay time of $\tau = hm$. When $\mathrm{sgn}_{\Delta u_i}= + 1$, the MCU outputs a positive control signal to charge the external circuit. Conversely, when $\mathrm{sgn}_{\Delta u_i}=-1$, the MCU initiates a discharging operation for the external circuit.

The width of the H-bridge drive pulse, denoted as $T_{\Delta u_i}$, is expressed as:
\begin{equation}
\begin{split}\label{e16}
T_{\Delta u_i}\propto\left| r_i x_i(n) \left( 1 - s_i x_i(n - m) \right)+\frac{\sum_{j = 1}^{N} a_{ij} x_i(n) x_j(n)}{D_i + E_i x_i(n)} \right|.
\end{split}
\end{equation}
The voltage signal output by the subtractor is given by:
\begin{equation}
\begin{split}\label{e17}
v_i(n)=V\cdot\mathrm{sgn}_{\Delta u_i}.
\end{split}
\end{equation}
where $V$ represents the amplitude of the control voltage. Assuming that the input resistance of the integrator is $R$ and the feedback capacitor is $C_1$, the voltage increment across the capacitor satisfies the inverse integration relationship:
\begin{equation}
\begin{split}\label{e18}
\Delta u_i&=-\frac{1}{RC_1}\int_{0}^{T_{\Delta u_i}}v_i\mathrm{d}t = -\frac{1}{RC_1}v_i(n)T_{\Delta u_i}\\
&=-\frac{1}{RC_1}V\cdot\mathrm{sgn}_{\Delta u_i}\cdot T_{\Delta u_i}\\
&\propto-\left[r_i x_i(n) \left( 1 - s_i x_i(n - m) \right)+\frac{\sum_{j = 1}^{N} a_{ij} x_i(n) x_j(n)}{D_i + E_i x_i(n)} \right]\\
&=-\Delta x_i.
\end{split}
\end{equation}
Consequently, through the acquisition of the output voltage signal using an oscilloscope and subsequent reverse deduction, the following relationship can be derived:
\begin{equation}
\begin{split}\label{e19}
\overline{u_i}=-u_i\propto x_i.
\end{split}
\end{equation}

This outcome successfully achieves the electrical mapping of the node state variable $x_i$. The dynamic response curve displayed on the oscilloscope can provide real-time and intuitive insights into the oscillation state of the network, thereby offering a reliable physical interface for experimental observations and theoretical validations.

\bibliographystyle{unsrt}
\bibliography{reference}

\end{document}